\documentclass{article}

\usepackage{graphicx}
\graphicspath{{Images}}
\usepackage{subcaption}
\usepackage{amsmath}
\usepackage{amssymb}
\usepackage{tabularx}
\usepackage{hyperref}

\newtheorem{definition}{Definition}
\newtheorem{assumption}{Assumption}
\title{\LARGE \bf
Improving Computational Cost of Bayesian Optimization for Controller Tuning with a Multi-stage Tuning Framework
}
\author{Marlon J. Ares-Milian $^{1}$, Gregory Provan $^{1}$, and Marcos Quinones-Grueiro $^{2}$% <-this % stops a space
\thanks{$^{1}$Marlon J. Ares-Milian and Gregory Provan are with the Department of Computer Science,
        University College Cork, Cork, Ireland
        {\tt\small 122100743@umail.com}, %
        {\tt\small gprovan@cs.ucc.ie}}
\thanks{$^{2}$Marcos Quinones-Grueiro is with the Department of Computer Science,
        Vanderbilt University, Tennessee, USA
        {\tt\small marcos.quinones.grueiro@Vanderbilt.Edu}}%
}

\begin{document}
\maketitle
\thispagestyle{empty}
\pagestyle{empty}

%%%%%%%%%%%%%%%%%%%%%%%%%%%%%%%%%%%%%%%%%%%%%%%%%%%%%%%%%%%%%%%%%%%%%%%%%%%%%%%%%%%%
\begin{abstract}
Control auto-tuning for industrial and robotic systems, when framed as an optimization problem, provides an excellent means to tune these systems. However, most optimization methods are computationally costly, and this is problematic for high-dimension control parameter spaces. In this paper, we present a multi-stage control tuning framework that decomposes control tuning into subtasks, each with a reduced-dimension search space. We show formally that this framework reduces the sample complexity of the control-tuning task. We empirically validate this result by applying a Bayesian optimization approach to tuning  multiple PID controllers in an unmanned underwater vehicle benchmark system. We demonstrate an 86\%  decrease in computational time and 36\% decrease in sample complexity.
%, a sample efficient optimization solution that has become popular in control auto-tuning but that scales poorly with search space dimension. We validate our results  with. % where a 30\% computational time improvement is achieved].
\end{abstract}

% Keywords
%\keyword{bayesian optimization; control auto-tuning} 

%%%%%%%%%%%%%%%%%%%%%%%%%%%%%%%%%%%%%%%%%%%%%%%%%%%%%%%%%%%%%%%%%%%%%%%%%%%%%%%%%%%%
\section{Introduction}

Control engineering is a core component of the operation of industrial and robotic systems. Proper design, implementation, and maintenance of control systems enables systems to meet a variety of performance objectives such as safe operation, robustness, low environmental impact, low energy consumption, high profit margins, etc.

Control algorithms (or controllers) are characterized by  parameters which must be adjusted, or tuned, to meet the desired performance objectives. For example, a PID controller has P, I and D parameters.
Traditional controller fine-tuning is done by hand, which is quite inefficient; this approach requires some degree of expert knowledge, and can only achieve sub-optimal solutions due to the high number of performance evaluations needed.

% Control tuning methodologies are very diverse, some require a high degree of plant(system)-model knowledge, for instance, linear quadratic regulators (LQR) or some versions of model predictive control (MPC)

Limitations in manual control tuning have motivated the introduction of various degrees of automation.
%Modern control tuning methodologies introduce various degrees of automation. 
For example, this may include an iterative experimental approach, in which performance objectives are evaluated (either in simulation or hardware) and the \textit{best} set of parameters is selected. 
%However, these performance evaluations are usually very costly. Alternative approaches  follow fixed tuning rules \cite{Cheng2023}\cite{Somefun2021}\cite{Paulson2023}, and may require no experiments: for instance, linear quadratic Gaussian control (LQG), H-infinity control, and quantitative feedback theory (QFT). However, fixed approaches usually require a final fine-tuning process on the real system to address potential uncertainties, which usually also results in an iterative trial-and-error approach. 
%---------------------------------
More recently, several automated tuning (auto-tuning) methodologies have combined system-model knowledge, expert knowledge, historical data, and software tools. % to determine the set of control parameters that optimizes the desired performance objectives. 
%Control auto-tuning has been a topic of research for proportional-integral-derivative (PID) controllers \cite{Cheng2023}, which is the most widely used control algorithm today \cite{Somefun2021}\cite{Mappas2023}. 
Relevant control auto-tuning research in recent years includes the use of reinforcement learning (RL) \cite{Mate2023}\cite{McClement2022}, evolutionary algorithms \cite{Joseph2022}\cite{RodriguezMolina2020}, and Bayesian Optimization \cite{Stenger2022bench}\cite{Stenger2022joint}, among others. Solving the control auto-tuning optimization problem is hard in general due to the costly black-box nature of its objectives. 

Bayesian Optimization (BO) auto-tuning, in particular, has gained traction in recent studies \cite{Rohr2024}, due to its sample-efficiency \cite{Stenger2022bench}, (which results in fewer system evaluations), and the fact that it is a model-free approach (i.e., does not require system-model knowledge) \cite{Nitsch2023}. However, BO is limited to optimization problems with low-dimensional search spaces \cite{Paulson2023}. This makes it feasible only in local parameter search for control algorithms with a small number of parameters. This is particularly challenging in systems with multiple inputs and multiple outputs (MIMO), where the number of control parameters can be high.
%Bayesian Optimization has become very popular for its theoretical guarantees [cite]

Most industrial and robotic systems have multiple inputs and multiple outputs (MIMO) \cite{Somefun2021} which, in turn, often leads to multiple control loops with independent controllers. Many tuning methods simplify the MIMO requirement  \cite{Lawrence2022}\cite{Kumar2021}, since tuning multiple controllers is significantly harder than tuning a single controller due to the increased number of parameters and coupling/interactions between variables \cite{Mate2023}. The high dimensionality of the control parameter space, coupled with the costly black-box nature of the system evaluations, makes control auto-tuning for MIMO systems computationally expensive,
%sometimes intractable, or at least, very costly from a computational point of view, 
both in terms of computation time and sample efficiency. Motivated by the challenges in MIMO control tuning, several works have tried to decompose the system into subsets of control loops. A common strategy consists of identifying control loops in the system that have few or no interactions with each other, which enables the tuning of each of these isolated control loops independently \cite{Pandey2022}. This is often achieved through expert knowledge \cite{Sahrir2022}; however, research on variable decoupling and control loop isolation is still an open issue \cite{Liu2019survey}.

%A motivation exists to reduce the computational complexity of the auto-tuning task; for example, to reduce the fuel, or raw materials costs of running tuning experiments on real hardware, or to quickly re-tune controllers in some adaptive control scenario. 
%However, most control auto-tuning solutions do not scale %(in terms of computational complexity) 
%with the dimensionality of the control parameter space. This is a widely acknowledged challenge in recent works on Bayesian Optimization for control auto-tuning \cite{Stenger2022bench}\cite{Paulson2023}; however, to the best of our knowledge, solutions for this problem have not been proposed. 

Our contributions are as follows.
\begin{enumerate}
\item 
To address the challenges in auto-tuning MIMO systems with high-dimension parameter spaces, we propose a multistage, optimisation-based control tuning framework for MIMO systems with decentralized control, which employs control loop decoupling and isolation. Our proposed framework is the first, to the best of our knowledge, to formalize control tuning for multiple control loops as a multistage optimization task. 
%----------------
\item 
We apply this multistage tuning framework to redefine the control tuning task into a set of tuning optimisation subtasks, and show that decomposing the control parameter search space into smaller-dimensional search spaces at each stage results in improved computational complexity for the auto-tuning solution task.
%---------------
\item We formally show this computational complexity improvement using Bayesian Optimization, and empirically validate it by solving the control auto-tuning problem for an autonomous underwater vehicle (AUV) with multiple PID control loops. We demonstrate an 86\%  improvement in computational time and 36\% decrease in sample complexity.
\end{enumerate}

This paper is structured as follows: Section \ref{sec::related_works} discusses the state-of-the-art research on control auto-tuning with an emphasis on Bayesian Optimization approaches. Section \ref{sec::preliminaries} formalizes the main concepts underpinning our approach. Section \ref{sec::tuning_methodology} presents the proposed multi-stage control tuning framework. Finally, Section \ref{sec::experimental_design} describes the experimental design to empirically validate our hypothesis, and in Section \ref{sec::results} we present and discuss our experimental results.

%%%%%%%%%%%%%%%%%%%%%%%%%%%%%%%%%%%%%%%%%%
\section{Related Works}\label{sec::related_works}
We are interested in automated controller tuning methods (or control auto-tuning methods) framed as an optimization problem. Within this framework, we can group existing auto-tuning approaches within two groups \cite{Agrawal2020}\cite{Paulson2023}: gradient-based and derivative free optimization (DFO) methods.

First-order (also referred to as gradient-based) optimization methods have been commonly used in control tuning \cite{Agrawal2020}. These methods rely on knowing the gradients of the performance objectives with respect to the control parameters, which can be achieved with previous knowledge of the system model \cite{Trimpe2014}\cite{Kumar2021} (model-based approaches) or using some form of gradient estimation \cite{Carlucho2020}\cite{Mate2023} (model-free approaches). Information on the gradients is then used to guide the search for the optimal control parameters. Reinforcement-learning (RL) is a well-known gradient-based learning algorithm that has seen use in control auto-tuning. Authors in \cite{Lawrence2022} propose an actor-critic RL solution where the actor is a PID controller function that is tuned following a constrained RL tuning algorithm. In \cite{Mate2023} multiple SISO PIDs are tuned simultaneously for a MIMO system using a single RL policy tuned with the deep deterministic policy gradient (DDPG) algorithm. A meta-reinforcement learning algorithm is proposed in \cite{McClement2022} to adaptively tune PI controllers when the underlying system dynamics change or are not contained in the training distribution. Overall, gradient-based approaches are limited by the amount of knowledge available about the system, or the accuracy of the gradient estimates \cite{Cheng2023}. Furthermore, they are not easily extended to consider optimization constraints \cite{Paulson2023}.

Derivative-free optimization methods (DFO) have become increasingly popular since they make little-to-no assumptions on the objective or constraint functions. Examples of DFO control tuning methods are Bayesian Optimization \cite{Stenger2022bench}, DiffTune \cite{Cheng2023} and evolutionary optimization \cite{Chang2014}\cite{Joseph2022}. 

Meta-heuristics-based control tuning algorithms are a well known example of DFO methods for control tuning; authors in \cite{RodriguezMolina2020} provide a survey on multi-objective optimization (MOO) using meta-heuristics, while \cite{Joseph2022} survey the particular case of PID controller tuning with meta-heuristics. These approaches include the use of grey wolf optimization \cite{Yadav2020} for PID controller tuning, non-dominated sorting Genetic Algorithm II (NSGA-II) for multi-objective tuning of aerial manipulators \cite{Zhou2019moo}, particle swarm optimization for PID tuning of MIMO processes \cite{Chang2014}, and chaotic atom search for tuning of fractional order PID controllers for DC motor speed control \cite{Hekimoglu2019}. 

As an alternative to meta-heuristics-based methods, Bayesian Optimization is a class of machine learning solutions designed for noisy optimization problems with expensive objective evaluations \cite{Paulson2023} that has become very popular for controller tuning in recent works \cite{Mulders2020}\cite{Kavas2024}. The main appeal in BO comes from its sample efficiency, shown by \cite{Stenger2022bench} to outperform a range of state-of-the-art DFO approaches for control auto-tuning, including several evolutionary algorithms. %BO has been shown to tune multiple families of control algorithms such as MPC [Paulson2023][cite], LQR controllers [Paulson2023][Stenger2022joint], PID controllers []

Authors in \cite{Stenger2022joint} present a joint tuning methodology which uses Bayesian Optimization to simultaneously tune an LQR controller, an unscented Kalman Filter (UKF), and a guidance and navigation system for an autonomous underwater vehicle (AUV), yielding satisfactory results across multiple objectives. They address the issue of over-fitting by tuning on a variety of different scenarios, which significantly increases the computational cost of evaluating the objective functions. Furthermore, the combination of multiple objectives (and potential cost parameter tuning) is not explicitly defined for the proposed joint approach. An earlier version of this paper is presented by \cite{Nitsch2023} where only the Kalman filters are tuned for the AUV using Bayesian Optimization.

In their work, \cite{Stenger2022bench}, \cite{Stenger2023thesis} and \cite{Rohr2024} propose a BO-based tuning methodology for learning with crash constraints (LCC) using virtual data-points. These works show BO to be very sample efficient for control tuning tasks (particularly when learning with crash constraints) in comparison to other state-of-the-arts auto-tuning optimizers.

A general disadvantage of DFO methods is the fact that they do not scale effectively with the increase in the dimension of the parameter space \cite{Paulson2023}. Furthermore, some of these methods converge very slowly compared to their gradient-based counterparts \cite{Agrawal2020}.

%\paragraph{MIMO Systems and Variable Decoupling for Control Tuning}

%%%%%%%%%%%%%%%%%%%%%%%%%%%%%%%%%%%%%%%%%%
\section{Preliminaries}\label{sec::preliminaries}
In this section, we formalize some key concepts, relevant to the rest of the paper.
\subsection{System definition.}\label{sec::system}
\begin{definition}[Non-linear Continuous System]
We define a non-linear continuous system with a tuple $S = (\mathcal{X}, \mathcal{U}, \mathcal{Y}, \mathcal{T}, f, U)$ where:
\end{definition}

\begin{itemize}

\item $\mathcal{X} \subseteq \mathbb{R}^{n_x}$: is the space of state variables.
\item $\mathcal{U} \subseteq \mathbb{R}^{n_u}$: is the space of control outputs (or system inputs).
\item $\mathcal{Y} \subset \mathbb{R}^{n_y}$: is a measurement space such that $n_y \leq n_x$. 
\item $\mathcal{T} \in \mathbb{R}_+$: is a time set characterized by variable $t\in \mathbb{R}_+$
\item $f:\mathcal{X} \times \mathcal{U} \to \mathcal{X}$: is a mapping that describes system dynamics such that:
\begin{equation}\label{eq::system_model_cont}
\dot{\mathbf{x}}(t) = f(\mathbf{x}(t), \mathbf{u}(t)),
\end{equation}

where $\mathbf{x}(t) \in \mathcal{X}$ and $\mathbf{u}(t) \in \mathcal{U}$.
\item $U:\mathcal{X}^{t} \times \Xi \to \mathcal{U}$: is a control law, parameterized by $\boldsymbol\xi \in \Xi$, that potentially includes information over time window $t$ such that:

\begin{equation}
\mathbf{u}(t) = U(\mathbf{x}(0),\dots, \mathbf{x}(t), \boldsymbol\xi)
\end{equation}

\end{itemize}

\begin{assumption}
We assume a subset of the state variables are fully measurable, i.e.: $\mathbf{y}(t) = C \mathbf{x}(t), \forall t \in \mathcal{T}, \mathbf{y}(t) \in \mathcal{Y}, \mathbf{x}(t) \in \mathcal{X}$; where $C \in M_{n_y \times n_x}$ is a binary matrix with exactly one entry of 1 in each row, and at most one entry of 1 in each column, with all other entries 0.
\end{assumption}

\subsection{Bayesian Optimization}\label{sec::bayesopt_def}
One can restate the problem of tuning the controller parameters as optimizing a  metric that guarantees appropriate system performance in its closed-loop operation. 

Bayesian optimization is a technique for globally optimizing black-box functions that are expensive to evaluate \cite{Jones1998}\cite{Kushner1964}\cite{Mockus1975}. In our setting, we consider the global minimization problem

\begin{equation}
\begin{aligned} 
{\boldsymbol{\xi}}^{*} = \underset{{\boldsymbol{\xi}}\in {\Xi}}{\arg \min }\, f_{\xi}({\boldsymbol{\xi}}), 
\end{aligned}
\end{equation}

with input space ${\Xi}=[0,1]^{D}$ and objective function $f_{\xi}:{\Xi}\rightarrow {\mathbb {R}}$. We consider functions $f_{\xi}$ that are costly to evaluate and for which we are allowed a small budget of evaluation queries to express our best guess of the optimum’s location 
${\boldsymbol{\xi}}^{*}$ in at most $T_{end}$ iterations. We further assume we have access only to noisy evaluations of the objective $l=f_{\xi}+\varepsilon$, where 
$\varepsilon \sim {\mathcal {N}}(0, \sigma _{n}^{2})$ is i.i.d. Gaussian measurement noise with variance 
$\sigma _n^2$. We restrict ourselves to the typical setting, where neither gradients nor convexity properties of $f_{\xi}$  are available.

The main steps of a BO routine at iteration $t$ involve (1) response surface learning, (2) optimal input selection ${\boldsymbol{\xi}}_{t+1}$ and (3) evaluation of the objective function $f_{\Xi}$ at ${\boldsymbol{\xi}}_{t+1}$. 

The BO framework uses the predictive mean and variance of the GP model to formulate an acquisition function that trades off exploitation, testing promising controller parameters given the current knowledge, with exploration, sampling unexplored regions of the design space.
 
The BO algorithm can be initiated without any past observations, but it is usually more efficient to calibrate the GP model hyper-parameters and calculate a prior by first collecting an initial set of observations. 

We can consider almost arbitrary performance metrics and safety indicators. However, here we only consider a metric representing a form of trajectory tracking error.

%%%%%%%%%%%%%%%%%%%%%%%%%%%%%%%%%%%%%%%%%%%%%%%%%%%%%%%%%%%

\subsubsection{Gaussian Processes}

A Gaussian process (GP) \cite{Rasmussen2006} is a distribution over the space of functions commonly used in non-parametric Bayesian regression. It is fully described by a mean function $\mu :\Xi \to \mathbb{R}$, which, WLOG, we set to zero for all inputs $\mu (\xi) = 0,\quad \forall \xi \in \Xi$, and a kernel function $k:\Xi \times \Xi \to \mathbb{R}$. Given the data set $\mathcal{D} = \left\{ {({\xi_i},{l_i}\} _{i = 1}^t} \right.$, where
$l_i = f(\xi_i) + \epsilon_i$ and $\epsilon_i \sim N(0,\sigma^2)$ is zero-mean i.i.d. Gaussian noise, the posterior belief over the function $f$ has the following mean, variance and covariance:

\begin{align*} 
& {\mu _t}(\xi) = {\mathbf{k}}_t^ \top (\xi){\left( {{{\mathbf{K}}_t} + {\sigma ^2}{\mathbf{I}}} \right)^{ - 1}}{{\mathbf{l}}_t},\tag{2} \\ & {k_t}\left( {\xi,\xi'} \right) = k\left( {\xi,\xi'} \right) - {\mathbf{k}}_t^ \top (\xi){\left( {{{\mathbf{K}}_t} + {\sigma ^2}{\mathbf{I}}} \right)^{ - 1}}{{\mathbf{k}}_t}\left( {\xi'} \right),\tag{3} \\ & {\sigma _t}(\xi) = {k_t}(\xi,\xi),
\end{align*}

where
$\mathbf{k}_t (\xi) =(k(\xi_1,\xi),\cdots k(\xi_t,\xi)),~ \mathbf{K}_t$ is the positive definite kernel matrix ${\left[ {k\left( {\xi,\xi'} \right)} \right]_{\xi,\xi' \in {\mathcal{D}_t}}},$, and $I\in \mathbb{R}^{t\times t}$  denotes the identity matrix. In the following, the superscripts $f$ and $q$ denote GPs on the objective and on the constraints.
 
%-----------------------------------------------------------
 \subsubsection{Computational Complexity}

 A BO routine consists of two key steps: (1) estimating the response surface, which is a black-box function from data through a probabilistic surrogate model, usually a Gaussian process(GP); (2) maximizing an acquisition function that trades off exploration and exploitation according to uncertainty and optimality of the response surface. 
The computational complexity of these two aspects are as follows:
%-----------------
\begin{itemize}
\item  The computational complexity of training a GP is typically $O(n^3)$,  where $n$  is the number of observations. This is due to the need to invert an $n \times n$  covariance matrix.
\item  The complexity of optimizing the acquisition function can range from $O(d)$  to $O(d^2)$  per iteration, where $d$  is the dimensionality of the search space, depending on the method used. Common methods include gradient-based optimization and evolutionary algorithms.  
\end{itemize}

For real-world applications where we have tens of thousands of observations and large-dimensional problems, this approach is computationally infeasible.
As the dimensionality of the input space increases, these two steps become intractable. From a machine learning or estimation perspective, the sample complexity to ensure good coverage of inputs for learning the response surface is exponential in the number of dimensions \cite{Shahriari2016}. Under standard assumptions (e.g., a small evaluation budget), the learned response surface and the resulting acquisition function are characterized by vast flat regions interspersed with highly non-convex landscapes \cite{Rana2017}, which makes maximizing a high-dimensional acquisition function to be inherently hard \cite{Garnett2014}.

In our case, we address  reducing the sample complexity  for learning the response surface, which is exponential in the number of dimensions. In other words, by reducing the number of dimensions (e.g., from $D$ to $D'$), we reduce the sample complexity by an exponential margin, e.g., by $O(2^{D-D'})$.

%%%%%%%%%%%%%%%%%%%%%%%%%%%%%%%%%%%%%%%%%%
\section{Control Tuning Framework}\label{sec::tuning_methodology}
We define the control tuning task as a multi-objective optimization (MOO) problem:

\begin{equation}\label{eq::moo}
\begin{aligned}
\min_{\boldsymbol{\xi}} \quad & \mathbf{L}(\boldsymbol{\xi}) = [\mathcal{L}_1(\boldsymbol{\xi}), \dots, \mathcal{L}_{n_L}(\boldsymbol{\xi})]\\
\textrm{s.t.} \quad & \boldsymbol{\xi}\in \Xi \\
\end{aligned}
\end{equation}
where $\boldsymbol{\xi} \in \Re^{n_\xi}$ is a vector of control parameters and $\mathbf{L}(\boldsymbol{\xi})$ is a vector of objectives. The proposed framework makes no assumptions on the parameters and objective spaces.

\subsection{Control tuning as a multistage problem.}\label{sec::multistage_framework}
\begin{figure}[h!]
\centering
\includegraphics[width=0.4\textwidth]{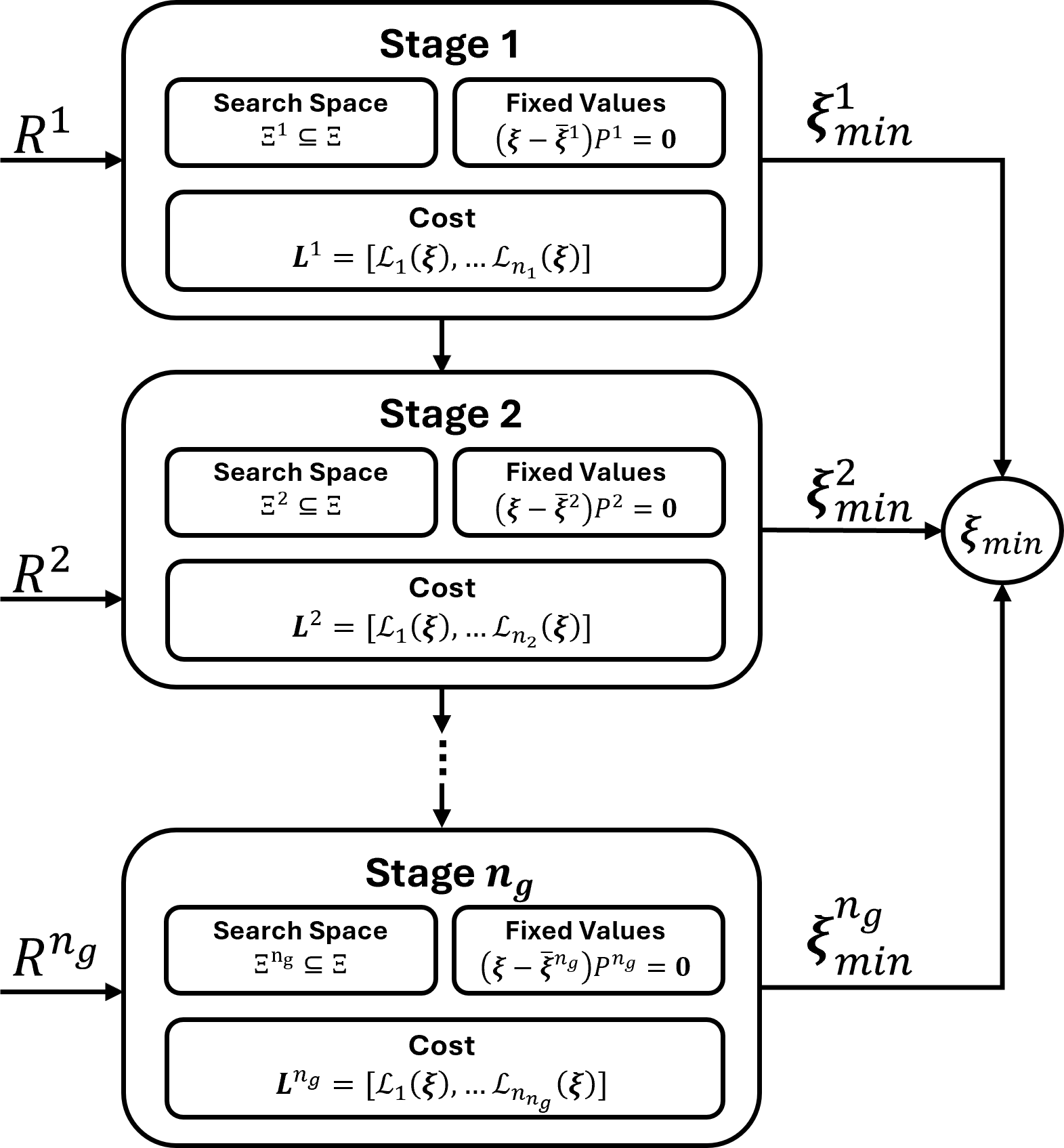}
\caption{Control Tuning as a Multistage Optimization Problem}
\label{fig::multistage_opt}
\end{figure}

This general tuning problem can be redefined as a multistage optimization problem where a subset of controllers is tuned at every stage of the pipeline. Each stage can be defined as a MOO subtask. Figure~\ref{fig::multistage_opt} illustrates the process.

At each stage ($i = 1, 2, \dots, n_g$) of the tuning process, a subset of control parameters ($\Xi^i \subseteq \Xi$) are tuned to minimize a set of stage-specific objectives ($\mathbf{L}^i$). The objectives are optimized given a reference signal $R^i$. The rest of the control parameters are fixed at a previous stage or before the tuning process begins; this is done through the constraint parameters $(\bar{\boldsymbol\xi}^i, P^i)$, where $P^i \in diag(\{0, 1\}^{n_\xi})$. At each stage, a subset ($\boldsymbol\xi^i_{min}$) of the optimal control parameters ($\boldsymbol\xi_{min}$) is produced as a result of a (potentially) multi-objective optimization subtask described in Section \ref{sec::moo_subtask}. 

The number of stages ($n_g \leq n_u$) of the tuning process is task-specific, and depends on the assumptions made regarding the degree of input-output decoupling of the system; e.g.: $n_g = 1$ for the case of a fully coupled system, $n_g = n_u$ for the case of a fully decoupled system. A variety of variable decoupling methodologies \cite{Liu2019survey}\cite{Pandey2022} can be considered in order to define the stages in the framework. However, for the purpose of this paper, we will define the stages using expert knowledge. In future works we will perform a more detailed analysis on the implications of these decoupling assumptions and their impact on control tuning performance.

%and considering the proposed control tuning framework in this paper is independent from the performance metric used (as long as it does not violate the continuity assumption for the GP in BO),

\subsection{Multi-objective optimization subtasks.}\label{sec::moo_subtask}
\begin{figure}[h!]
\centering
\includegraphics[width=0.4\textwidth]{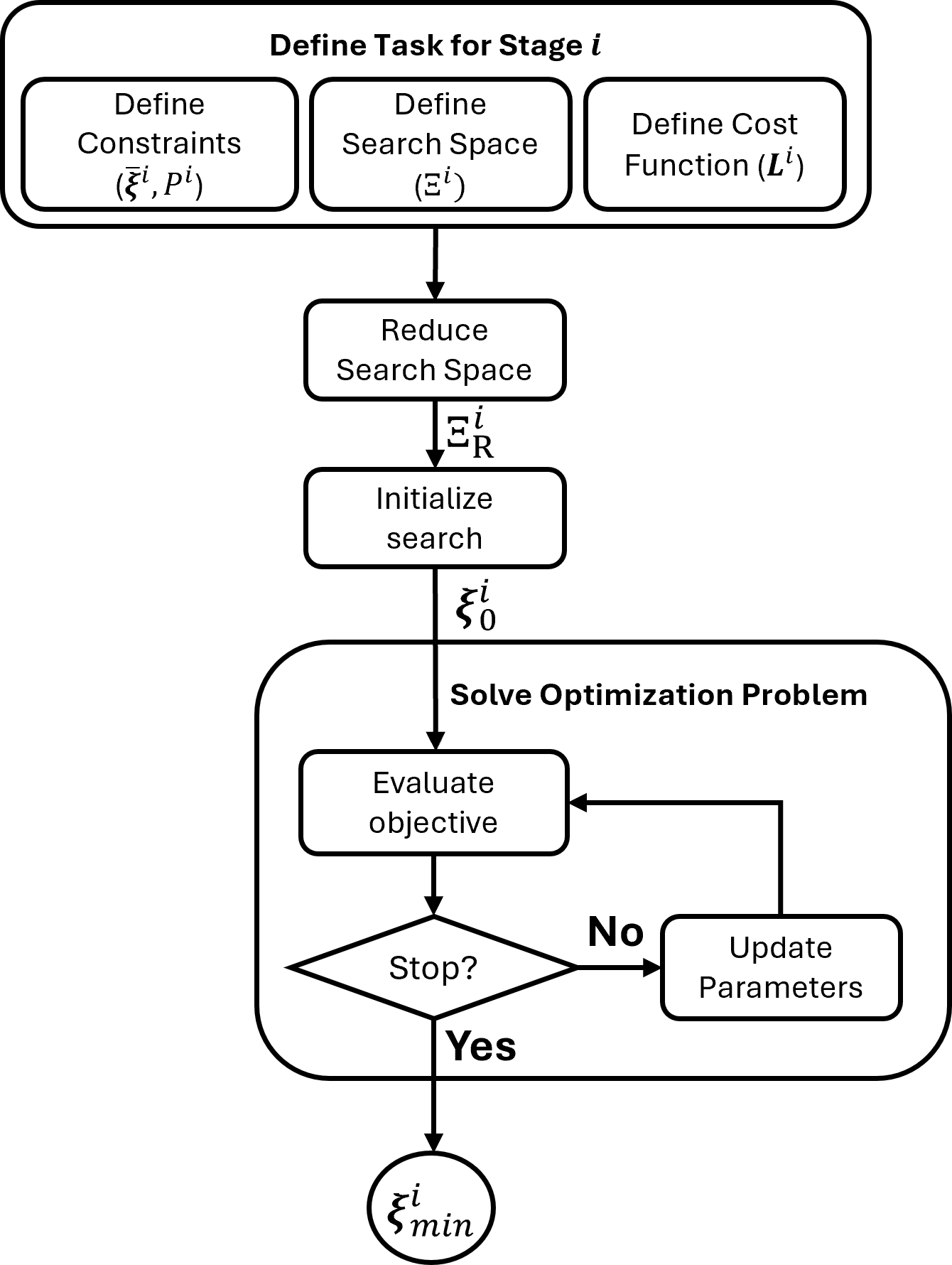}
\caption{Control Tuning Multi-Objective Optimization Subtask}
\label{fig::opt_subtask}
\end{figure}

The control-tuning optimization task is redefined into optimization subtasks at every stage (Figure~\ref{fig::opt_subtask}). Each stage $i$ can be separated into 3 main steps: 
\begin{enumerate}
    \item An initial step where constraints, search space, and objectives are defined.
    \item A space restriction step where the search space may be restricted to $\Xi^i_R \subseteq \Xi^i$ and a parameter initialization $\boldsymbol{\xi}^i_0$ may be proposed. The purpose of this step is to potentially simplify the parameter search for every subtask; examples of this could be the application of Ziegler-Nichols tuning rules \cite{Somefun2021} before fine-tuning of a controller, or the restriction of the controller space in order to achieve some theoretical guarantees.
    \item The parameter search step, where the multi-objective optimization subtask is solved using any methodology available and the optimal parameters $\boldsymbol{\xi}^i_{min}$ are produced. Examples of methodologies could be: manual tuning, meta-heuristics, gradient-based solutions, etc.
\end{enumerate}
The multi-objective optimization subtask is defined as follows:
\begin{subequations}
\label{eq::moo_subtask}
\begin{align}
\min_{\boldsymbol{\xi}} \quad & \mathbf{L}^{i}(\boldsymbol{\xi}) = [\mathcal{L}_1(\boldsymbol{\xi}), \dots, \mathcal{L}_{n_i}(\boldsymbol{\xi})]\\
\textrm{s.t.} \quad & \boldsymbol{\xi}\in \Xi^i_R \\
& (\boldsymbol{\xi} - \bar{\boldsymbol{\xi}}^i)P^i = 0 \label{eq::moo_subtask_fix_par} 
\end{align}
\end{subequations}
where (\ref{eq::moo_subtask_fix_par}) is a constraint that sets control parameters to specified values. $P^i$ is an $n_\xi \times n_\xi$ binary diagonal matrix such that $p_{jj} \in P$ is 1 if the control parameter $j$ is fixed and 0 if its free; parameters are fixed to the values in vector $\bar{\boldsymbol{\xi}}^i$.

%\paragraph{Subtask multi-objective optimization pseudo-code}

\subsection{Bayesian Optimization for Optimization Subtasks}
The proposed tuning methodology presented in Section \ref{sec::multistage_framework} can be compatible with any optimization algorithm. For the purpose of this work, we are interested in its implementation using Bayesian Optimization to solve each optimization subtask as presented in (\ref{eq::moo_subtask}).

Following the definition in Section \ref{sec::bayesopt_def}, we want to solve the potentially local minimization problem for each optimization subtask $i$:

\begin{equation}
\boldsymbol{\xi}^{i}_{min} = \underset{{\boldsymbol{\xi}}\in {\Xi_R^i}}{\arg \min }\, \mathcal{L}^i({\boldsymbol{\xi}}) 
\end{equation}
where $\boldsymbol{\xi}^{i}_{min}$ is the optimal control parameter configuration for the subtask, $\Xi_R^i$ is the input space, and $\mathcal{L}^i({\boldsymbol{\xi}}) = f(\mathcal{L}_1(\boldsymbol{\xi}), \allowbreak \dots, \mathcal{L}_{n_i}(\boldsymbol{\xi}))$ is some combination of potentially multiple objectives into a scalar value. Bayesian optimization is compatible with any performance metric as long as it does not violate the continuity assumption for the GP in BO.

The BO solution then consists in iteratively learning a Gaussian Process to map the control parameter space $\Xi_R^i$ to the performance objective $\mathcal{L}^i({\boldsymbol{\xi}})$ given a specific task (e.g.: path following, reference tracking, disturbance rejection, etc.). At each iteration, the task is evaluated and a new parameter-cost pair is produced, the mapping for the new pair is learned, and the BO acquisition function defines the next parameter configuration to evaluate. If available, $\boldsymbol{\xi}^i_0$ can be used to initialize the BO algorithm.

In practice, constraint (\ref{eq::moo_subtask_fix_par}) can be guaranteed by solving the optimization problem for a reduced vector of non-fixed control parameter values $\hat{\boldsymbol{\xi}}^i = \{\xi^i_j \in \boldsymbol{\xi}^i | p_{jj}^i = 0, p_{jj}^i \in P^i\}$ while fixing the rest of the parameters to their values in $\bar{\boldsymbol{\xi}}^i$.
\subsection{Improving Computational Cost}
In this section, we describe the potential computational cost improvements achieved by decomposing the search space using the proposed multistage framework. 

First, let's consider the computational complexity of solving the control tuning optimization using Bayesian Optimization for the complete search space. From Section \ref{sec::bayesopt_def}, we have a computational complexity for the full search space tuning of $O(2^{n_\xi})$, where ${n_\xi}$ (Section \ref{sec::multistage_framework}) is the number of control parameters in the system. 

A decomposition of the control tuning task using the multistage framework results in a set of $n_g$ lower dimension optimization tasks to be solved sequentially. This results in the following computational complexity: $\sum_{i=1}^{n_g} O(2^{n^i_\xi})$, where $n^i_\xi$ is the dimension of the search space for stage $i$, such that $n^i_\xi < n_\xi, \sum_{i=1}^{n_g} n^i_\xi = n_\xi$. 

We can clearly see that a decomposition of the search space using the proposed multistage framework results in  an improvement in computational complexity since $\sum_{i=1}^{n_g} O(2^{n^i_\xi}) < O(2^{n_\xi})$ is a linear combination of smaller dimension exponentially complex subtasks. Section \ref{sec::results_comp_cost} presents an empirical validation of this expected improvement in computational complexity for an underwater vehicle control auto-tuning benchmark.

%%%%%%%%%%%%%%%%%%%%%%%%%%%%%%%%%%%%%%%%%%
\section{Experimental Design. Underwater Vehicle Benchmark}\label{sec::experimental_design}
\subsection{Episode-based evaluations.}\label{sec::episode_sim}
In order to evaluate the cost function presented in Section \ref{sec::moo_subtask}, a run/simulation of the system must be performed. We define these runs to occur in episodes in which, for a fixed set of control parameters $\boldsymbol{\xi}$, the system is required to follow a reference signal while producing a sequence of inputs and outputs within a finite period of time. Formally:

\begin{definition}[Trajectory Tracking Episode]
\label{def::episode}
We define a trajectory tracking episode by a tuple $E = (\mathbf{T}, \mathbf{x}_0, R, \boldsymbol{\tau}, f^E)$ where:
\end{definition}
\begin{itemize}
\item $\mathbf{T} = \{0, T, 2T, \dots kT\}, k = 0, 1,\dots K$: is a vector of timestamps with a sampling period of $T$.
\item $\mathbf{x}_0$: is the state of the system at the start of the episode ($t = 0$).
\item $R:\mathcal{T} \to \mathbb{R}^{n_R}$, with $n_R \leq n_y$ is a reference signal; and $\mathbf{y}_R(kT) \in \{R(0), R(T),\allowbreak R(2T), \dots R(kT)\},  k = 0, 1,\dots K$ is a sampling of that signal.
\item $\boldsymbol{\tau} = \{(\mathbf{y}(0), \mathbf{u}(0)), \dots (\mathbf{y}(KT), \mathbf{u}(KT))\}$ is a temporally indexed sequence that is consistent with the underlying system model $f$ (defined in (\ref{eq::system_model_cont})). The vector $\boldsymbol{\tau}$ is the result of sampling the system inputs ($\mathbf{u}(t)$) and measurements ($\mathbf{y}(t)$) with a sampling period $T$ for the length of an episode ($KT$). We refer to $\boldsymbol{\tau}$ as the episode \textit{trail}.
\item $f^E$: is a black-box mapping of the system behavior during the episode such that $\boldsymbol{\tau} = f^E(\mathbf{x}_0, R, \mathbf{T}, \boldsymbol\xi)$. The resulting trail $\boldsymbol{\tau}$ is fully parameterized by $\boldsymbol\xi$ for a given initial state $\mathbf{x}_0$ and reference $R$. For the purpose of this work, we assume the system model has no uncertainties.
\end{itemize}
\subsection{Unmanned Underwater Vehicle (UUV)}
We evaluate the proposed control tuning methodology using an unmanned underwater vehicle benchmark (UUV) with six degrees of freedom (DOF), a MIMO system with complex dynamics. We use the Simu2VITA benchmark, a Matlab/Simulink\textsuperscript{\textregistered} simulator developed in \cite{Cerqueira2022}. This simulator provides a range of UUV modeling and simulation features including: vehicle dynamic and kinematic modeling, thruster dynamic simulation, control allocation, etc. More information regarding the UUV model can be found in \cite{Fossen2011} and \cite{Cerqueira2022}, and the full simulator can be found in our repository.

The control architecture for the vehicle consists of 6 independent proportional-integral-derivative (PID) feedback controllers. The control outputs (system inputs) for the vehicle are respectively summarized in $[F_x, F_y, F_z, T_\phi, T_\theta, T_\psi]$, with 3 PIDs controlling the forces in the $x$, $y$, $z$ axis and the other 3 PIDs controlling torques in roll $\phi$, pitch $\theta$, yaw $\psi$ orientation angles. The controllers respectively receive feedback from 6 measured variables: $[x, y, z, \phi, \theta, \psi]$ which measure the 3D position on the world frame and the 3 orientation angles.
%\subsection{UUV Control Algorithm}
% \begin{equation}
% \boldsymbol{\xi} = \{K_p^1, K_i^1, K_d^1, K_p^2, K_i^2, K_d^2, \dots, K_p^{n_u}, K_i^{n_u}, K_d^{n_u}\}
% \end{equation}
% and $\mathbf{K}^i = \{K_p^i, K_i^i, K_d^i\}$ represents the \textit{ith} PID controller. How to extend the proposed framework to non-PID parametric controllers is a topic for future works.
\subsection{Reference Tracking Task}
We evaluate the performance of our controller configurations for the UUV in terms of reference tracking tasks. We define two main tasks, a well-known step signal tracking task and a trajectory tracking task defined below. 

\subsubsection{Step tracking task}
Control tuning using a step reference signal is a very common practice even in complex systems \cite{Lawrence2022}\cite{Sahrir2022}.

A step signal $R_s(t):\mathcal T \to \Re $ is defined as:

\begin{equation}\label{eq::step_signal}
R_s(t) = 
    \begin{cases}
        R_0 & \text{if } t < t_e\\
      R_0 + R_A & \text{if } t \geq t_e
    \end{cases}
\end{equation}
where $R_A$ is the amplitude of the step signal, $R_0$ is the initial value of the step signal, and $t_e$ is the step start time. We assume that the initial state of the system ($x_0$) and the initial value of the step signal $R_0$ are the same for the purpose of this paper, i.e.: $x_0 = R_0$.
\subsubsection{Trajectory tracking task}\label{sec::traj_task}
We define the trajectory tracking task in terms of a matrix of reference way-points $\gamma \in \mathcal{M}_{n_w \times n_r}$, where $n_w$ is the number of way-points and $n_r$ is the dimension of the reference signal. The task consists of traversing these way-points sequentially until the last way-point is reached. A way-point is considered reached when the UUVs measured state is within a predefined radius ($w_r)$ of its reference, i.e.: $d(\mathbf{y}, \mathbf{y}_R) < w_r, \mathbf{y}_R \in \gamma$, where $d$ is some distance metric. Once a way-point $\mathbf{y}^j_R$ is reached, the next way-point in $\gamma$ becomes the reference $\mathbf{y}^{j+1}_R$; this continues until way-point $\mathbf{y}^{n_w}_R$ is reached and the trajectory is considered \textit{complete}, ending the episode (as defined in Section \ref{sec::episode_sim}). This task can be interpreted as a sequence of multidimensional step tracking tasks.
%The time $KT$ from the begining of the episode to the moment the trajectory is complete will be defined as the \textit{trajectory completion time}.

\subsection{Control Performance Metrics}
When solving an optimization problem, the design of an objective function can be a challenging task. In the case of control tuning, performance metrics are very diverse; they can include objectives regarding tracking error such as overshoot and steady state error, or time related objectives, like settling time, or trajectory completion time, as well as some metrics of control effort or energy consumption. Leveraging all of these objectives into a scalar cost function (or into some form of Pareto Front analysis \cite{RodriguezMolina2020}) can be very challenging, leading to higher experimental costs related to objective function parameterization or shaping. Probably for these reasons, a high degree of academic approaches to control tuning \cite{Chang2014}\cite{Cheng2023}\cite{Stenger2022joint}\cite{Rohr2024} usually employ some form of integral or mean error (e.g. root medium square error, integral absolute error), or time multiplied integral error as their scalar control performance objective. These functions are not significantly expressive or flexible when it comes to leveraging objectives, but they can integrate both temporal and tracking accuracy information into a single scalar value in a jack-of-all-trades manner. Their main advantage is their simplicity in implementation, hence their popularity in academic works where the main contributions are not related to (or are independent from) cost function design. Following this reasoning, we will be using two integral-error-based control performance metrics for our optimization problem.

\subsubsection{Step Tracking Performance}
For the step tracking task, we will be using integral absolute error (IAE) \cite{Chang2014} as the performance metric to optimize using Bayesian Optimization. 
% \begin{equation}
% \Upsilon_{IAE}(\boldsymbol\tau) = \int_0^{KT} |\mathbf{y}(t) - \mathbf{y}_R(t)| dt
% \end{equation}
The BO objective $\mathcal{L}^i(\boldsymbol\xi)$ is evaluated by simulating an episode of fixed length $KT$  where the reference of one of the UUV controllers is stimulated with a step signal of magnitude $R_A$, producing an episode trail $\boldsymbol\tau$ (Section \ref{sec::episode_sim}). The IAE is then calculated for the corresponding output variable from the resulting episode trail such that $\mathcal{L}^i(\boldsymbol\xi) = \Upsilon_{IAE}(\boldsymbol\tau)$.

\subsubsection{Trajectory Tracking Performance}
For the trajectory tracking task we define a variation of the IAE that includes an exponential penalty for time. Time-multiplied error metrics \cite{Chang2024} are commonly used when the error by the end of an episode (where the response to a change in the reference is supposed to have stabilized) should be penalized more severely than at the beginning, prioritizing steady state over transient state accuracy. In this case, we define an exponentially-time-multiplied integral absolute error (eTxIAE) metric with the purpose of heavily penalizing episode length:
\begin{equation}
\Upsilon_{eTxIAE}(\boldsymbol{\xi}) = \int_0^{KT} e^t|\mathbf{y}(t) - \mathbf{y}_R(t)| dt
\end{equation}
%\subsection{Variable Decoupling using the Relative Gain Array}
The BO objective $\mathcal{L}^i(\boldsymbol\xi)$ is evaluated by simulating an episode of maximum length $K_{max}T$ where the reference sequentially follows a way-point matrix $\gamma$ until the trajectory is completed (at time $KT \leq K_{max}T$) producing an episode trail $\boldsymbol\tau$ (Section \ref{sec::episode_sim}). The eTxIAE is then calculated from the resulting episode trail such that $\mathcal{L}^i(\boldsymbol\xi) = \Upsilon_{eTxIAE}(\boldsymbol\tau)$.

\subsection{Tuning Controllers for the UUV as a Multi-stage Optimization Problem}
\subsubsection{Tuning controllers simultaneously for trajectory tracking}\label{sec::tune_simult}
In this section, we will describe the methodology followed to tune all of the controllers in the UUV simultaneously as a single trajectory-tracking problem. Following the framework described in Section \ref{sec::multistage_framework}, we define the trajectory tracking control tuning problem as a single-stage ($n_g = 1$) problem with the parameters presented in Table \ref{tbl::mstage_path}.

\begin{table}[h!]
\caption{Multistage Optimization Parameters for Coupled Tuning of a Vehicle Benchmark}
\label{tbl::mstage_path}
\begin{center}
\begin{tabular} {c|c|c}
Stage & $P^i$ & $\bar{\xi}^i$ \\
\hline
1     & $\{\mathbf{0}, \mathbf{0}, \mathbf{0}, \mathbf{0}, \mathbf{0}, \mathbf{0}\}$             
      & $\{\mathbf{-}, \mathbf{-}, \mathbf{-}, \mathbf{-}, \mathbf{-}, \mathbf{-}\}$       
      \\
\end{tabular}
\end{center}
\end{table}

The tuning problem becomes a standard, single optimization task where all 18 control parameters ($\boldsymbol\xi \in \Re^{18}$) must be tuned simultaneously using a BO approach for a trajectory tracking task. For the purpose of this paper, we tune the controllers to complete a single square-shaped trajectory in the $XY$ plane defined by way-point matrix $\gamma \in \mathcal{M}_{8 \times 6}$ where the $z$ position and the attitude angles ($\phi$, $\theta$, $\psi$) must be kept at a value of 0 during the whole trajectory tracking task. In Section \ref{sec::results_performance} we can find a graphical representation of the target trajectory; we refer the reader to our repository for more details on the trajectory parameters (\url{https://github.com/mjares/BOpt_AutoTuning_ECC25.git}).

The assumption of local search is very common in control auto-tuning as an optimization problem \cite{Rohr2024}\cite{Cheng2023}, particularly for BO-based auto-tuning, due to its limitations regarding dimensionality. Unfortunately, how to define these local search spaces is a topic that is, either not discussed, or poorly discussed in most recent works on control auto-tuning \cite{Cheng2023}\cite{Stenger2022joint}\cite{Rohr2024}\cite{Mate2023}, where these restricted spaces are apparently defined arbitrarily, when there is probably some degree of expert knowledge involved. We acknowledge this is still an open issue in control auto-tuning, however, it falls outside the scope of this paper. 

We define a local search space $\Xi_R^1$ for our tuning approach by manually tuning the controllers following the methodology in \cite{Sahrir2022}, and then arbitrarily selecting max-min ranges around this manually tuned solution. The resulting search space for the optimization task is presented in Table \ref{tbl::ranges}:

\begin{table}[h]
\caption{Local Search Space for Control Parameters}
\label{tbl::ranges}
\begin{center}
\begin{tabular} {c|c|c}
Controller & $[K_p^{min}, K_i^{min}, K_d^{min}]$ & $[K_p^{max}, K_i^{max}, K_d^{max}]$ \\
\hline
roll ($\phi$)  & $[0, 0, 0]$ & $[5, 5, 5]$ \\          
pitch ($\theta$) & $[0, 0, 0]$ & $[5, 5, 5]$ \\
yaw ($\psi$) & $[0, 0, 0]$ & $[5, 5, 5]$ \\
$x$   & $[150, 0, 75]$ & $[250, 10, 150]$ \\
$y$      & $[150, 0, 75]$ & $[250, 10, 150]$ \\
$z$    & $[145, 0, 95]$ & $[155, 5, 105]$ \\
\end{tabular}
\end{center}
\end{table}

A maximum episode time was set to $K_{max}T = 100s$ for the whole tuning process, however, an episode can end earlier if the trajectory is completed before $100s$.
 
\subsubsection{Tuning controllers independently for step tracking}\label{sec::tune_indiv}
In this section, we propose a methodology meant to decompose the problem in Section \ref{sec::tune_simult} into multiple smaller dimension optimization subtasks where each of the UUV controllers are tuned individually as a step tracking task. A well-known approach \cite{Sahrir2022}\cite{Fossen2011} to independently tuning controllers in a UUV assumes that attitude control variables are fully decoupled ($u_\phi$, $u_\theta$, $u_\psi$) and position control variables are fully decoupled ($u_x$, $u_y$, $u_z$). However, some degree of coupling between the attitude and position variables is assumed. The problem can be framed as a multistage ($n_g = 6$) optimization problem, following the multistage framework defined in Section \ref{sec::multistage_framework},  with the parameters in Table \ref{tbl::mstage_hierarchical}:
\begin{table}[h!]
\caption{Multistage Optimization Parameters for Hierarchical Tuning of a Vehicle Benchmark}
\label{tbl::mstage_hierarchical}
\begin{center}
\begin{tabular} {c|c|c}
Stage & $P^i$ & $\bar{\xi}^i$ \\ %& $\mathbf{L}^i$ \\
\hline
1     & $\{\mathbf{1}, \mathbf{1}, \mathbf{1}, \mathbf{0}, \mathbf{1}, \mathbf{1}\}$             
      & $\{\mathbf{0}, \mathbf{0}, \mathbf{0}, \mathbf{-}, \mathbf{0}, \mathbf{0}\}$        
      \\
\hline
2     & $\{\mathbf{1}, \mathbf{1}, \mathbf{1}, \mathbf{1}, \mathbf{0}, \mathbf{1}\}$          
      & $\{\mathbf{0}, \mathbf{0}, \mathbf{0}, \mathbf{0}, \mathbf{-}, \mathbf{0}\}$           
      \\
\hline
3     & $\{\mathbf{1}, \mathbf{1}, \mathbf{1}, \mathbf{1}, \mathbf{1}, \mathbf{0}\}$            
      & $\{\mathbf{0}, \mathbf{0}, \mathbf{0}, \mathbf{0}, \mathbf{0}, \mathbf{-}\}$          
     \\
\hline
4     & $\{\mathbf{0}, \mathbf{1}, \mathbf{1}, \mathbf{1}, \mathbf{1}, \mathbf{1}\}$        
      & $\{\mathbf{-}, \mathbf{0}, \mathbf{0}, \mathbf{K}^\phi_{min}, \mathbf{K}^\theta_{min}, \mathbf{K}^\psi_{min}\}$            
     \\
\hline
5     & $\{\mathbf{1}, \mathbf{0}, \mathbf{1}, \mathbf{1}, \mathbf{1}, \mathbf{1}\}$                     
      & $\{\mathbf{0}, \mathbf{-}, \mathbf{0}, \mathbf{K}^\phi_{min}, \mathbf{K}^\theta_{min}, \mathbf{K}^\psi_{min}\}$            
      \\
\hline
6     & $\{\mathbf{1}, \mathbf{1}, \mathbf{0}, \mathbf{1}, \mathbf{1}, \mathbf{1}\}$                    
      & $\{\mathbf{0}, \mathbf{0}, \mathbf{-}, \mathbf{K}^\phi_{min}, \mathbf{K}^\theta_{min}, \mathbf{K}^\psi_{min}\}$            
      \\
\hline
\end{tabular}
\end{center}
\end{table}

Each stage defines an optimization subtask for each of the UUV controllers where a step tracking task must be optimized. The restricted parameter spaces $\Xi^i_R$ for each stage are the same as the ones defined in Table \ref{tbl::ranges}. The attitude controllers are stimulated with a step reference signal of $R_A = 5^{\circ}$ and the position controllers with an $R_A = 3m$ step signal. Episode length is fixed to $KT = 20s$ for the whole tuning process. A step-by-step methodology is presented below:
\begin{enumerate}
\item Set the control parameters for all 6 PID controllers to zero.
\item Stimulate the yaw controller with a reference step signal while the rest of the DOF have a reference of 0. 
\item Tune the yaw controller (Stage 1) as an optimization subtask using BO. We start with the yaw controller since we are interested in prioritizing movements in the horizontal ($XY$) plane.
\item Repeat steps 2 and 3 for the roll and pitch controllers (Stages 2 and 3). While each of these controllers is being tuned, the parameters for the rest of the controllers are set to zero.
\item Set the attitude controllers (roll, pitch, yaw) to their tuned parameters ($\mathbf{K}^\phi_{min}, \mathbf{K}^\theta_{min}, \mathbf{K}^\psi_{min}$).
\item Stimulate the $x$ input with a reference step signal while the rest of the DOF have a reference of $0$.
\item Tune the $x$ controller (Stage 4) as an optimization subtask using BO. For this process, the attitude controllers are set to their tuned parameters in steps 2 to 3 ($\mathbf{K}^\phi_{min}, \mathbf{K}^\theta_{min}, \mathbf{K}^\psi_{min}$), while the other position controllers have their parameters set to zero.
\item Repeat steps 6 and 7 for the $y$ and $z$ controllers (Stages 5 and 6).
\end{enumerate}

%%%%%%%%%%%%%%%%%%%%%%%%%%%%%%%%%%%%%%%%%%
\section{Results and Discussion}\label{sec::results}
In this section, we compare the controller tuning results between the simultaneous tuning approach and the multistage-individual tuning approach using Bayesian Optimization to solve the optimization sub-tasks. We compare both approaches in terms of reference tracking performance for a square trajectory tracking task, as well as trajectory completion time (Section \ref{sec::results_performance}). We also present an empirical comparison between the computational cost in terms of computation time and number of samples (Section \ref{sec::results_comp_cost}). Results for a manual tuning approach (proposed in \cite{Sahrir2022}) are also presented as a baseline comparison. 

Each optimization-based tuning approach is run 4 times and mean and standard deviation values are reported for performance and computational complexity comparisons. We stop a  run  when a predefined maximum number of iterations has been reached, or a predefined cost function threshold is achieved. The maximum number of iterations for the simultaneous tuning approach is 1000; for the individual controller tuning sub-tasks, maximum iterations are 200 for the attitude controllers and 100 for the position controllers. We use MATLAB \textit{bayesopt} method for Bayesian Optimization, with the \textit{expected-improvement-plus} acquisition function.

\subsection{Comparing Tracking Performance}\label{sec::results_performance}
\begin{figure}[thpb]
  \centering
  \includegraphics[width=0.8\linewidth]{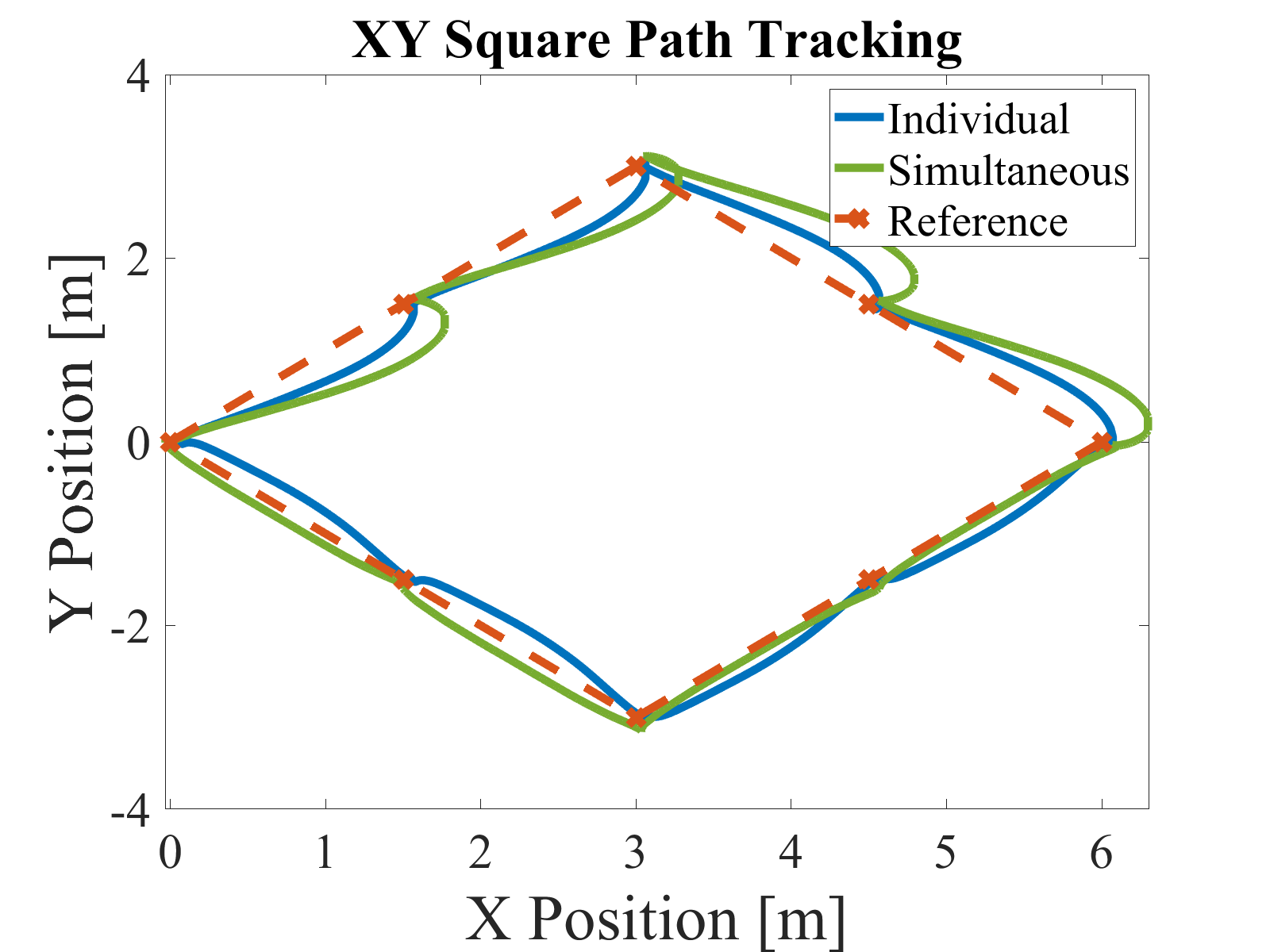}
  \caption{XY Plane Response}
  \label{fig::results_xy_path}
\end{figure}

Table \ref{tbl::results_performance} shows a comparison in performance (mean and standard deviation) between the manual tuning approach, the BO individual auto-tuning approach (Section \ref{sec::tune_indiv}), and the BO simultaneous auto-tuning approach (Section \ref{sec::tune_simult}). Tracking performance, in terms of $\Upsilon_{eTxIAE}$, and trajectory completion time (Section \ref{sec::traj_task}) are evaluated for the square trajectory presented in Fig. \ref{fig::results_xy_path}. The table shows that both auto-tuning approaches outperform the manually tuned controllers, which is an expected result. The table shows that the individual tuning approach outperforms the simultaneous tuning approach in both tracking error and completion time.

% We use the Wilcoxon signed-rank test \cite{Hollander2015} to compare the performance of the individual tuning-approach and the simultaneous tuning approach, since we cannot assume a normal distribution of the differences between the results from both approaches. The test for the tracking error shows a \textit{p-value} of under 0.266, rejecting, with a 5\% significance level, the hypotheses that the results from both tuning approaches come from a distribution with the same median. Similarly, the null hypotheses is also rejected for the completion time values (0.266).

\begin{table}[h] 
\caption{Performance Comparison for Tuning Approaches}
\newcolumntype{C}{>{\centering\arraybackslash}X}
\begin{tabularx}{\textwidth}{C|CCC}
\label{tbl::results_performance}
\textbf{Performance}	& \textbf{Manual}	& \textbf{BO Individual} & \textbf{BO Simultaneous}\\
% \hline
% Tracking error (IAE) & 1781.0 & 428.0 (52.54) & 365.6 (56.92)\\
\hline 
Tracking error (eTxIAE)	& 2.413e+68 	& 5.50e+21 (6.32e+21)  & 5.33e+34 (1.07e+35) \\  %\textsuperscript{1}\\
\hline
%Completion 
Time (s) & 158.9  & 48.8 (4.35) & 66.7 (12.89)\\
\end{tabularx}
\end{table}

\begin{figure}[thpb] 
\centering
\begin{subfigure}[t]{.8\textwidth}
  \centering
  \includegraphics[width=\linewidth]{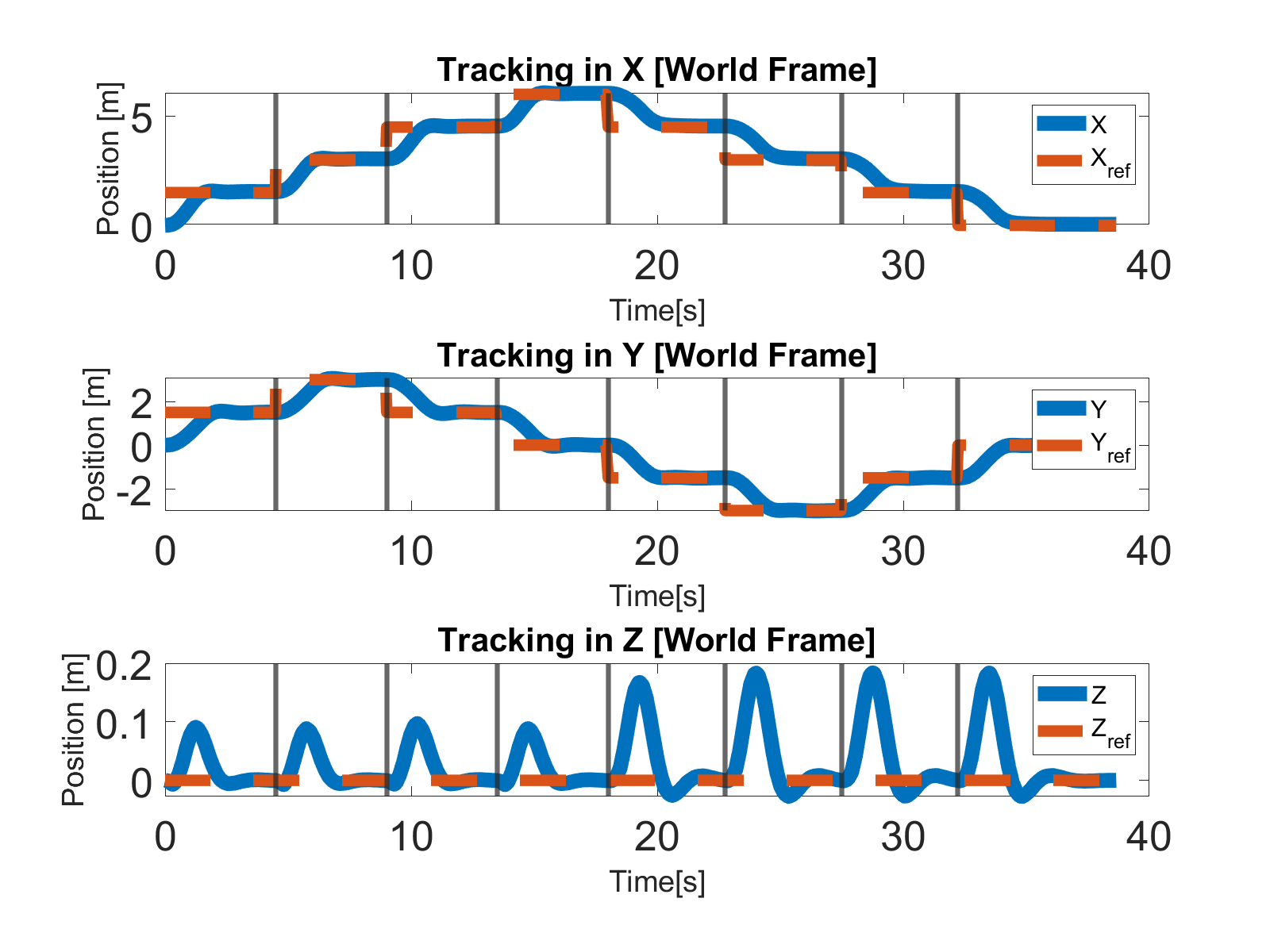}
  \caption{Position Variables Response}
  \label{fig::results_ind_tuned_pos}
\end{subfigure}
\begin{subfigure}[t]{.8\textwidth}
  \centering
  \includegraphics[width=\linewidth]{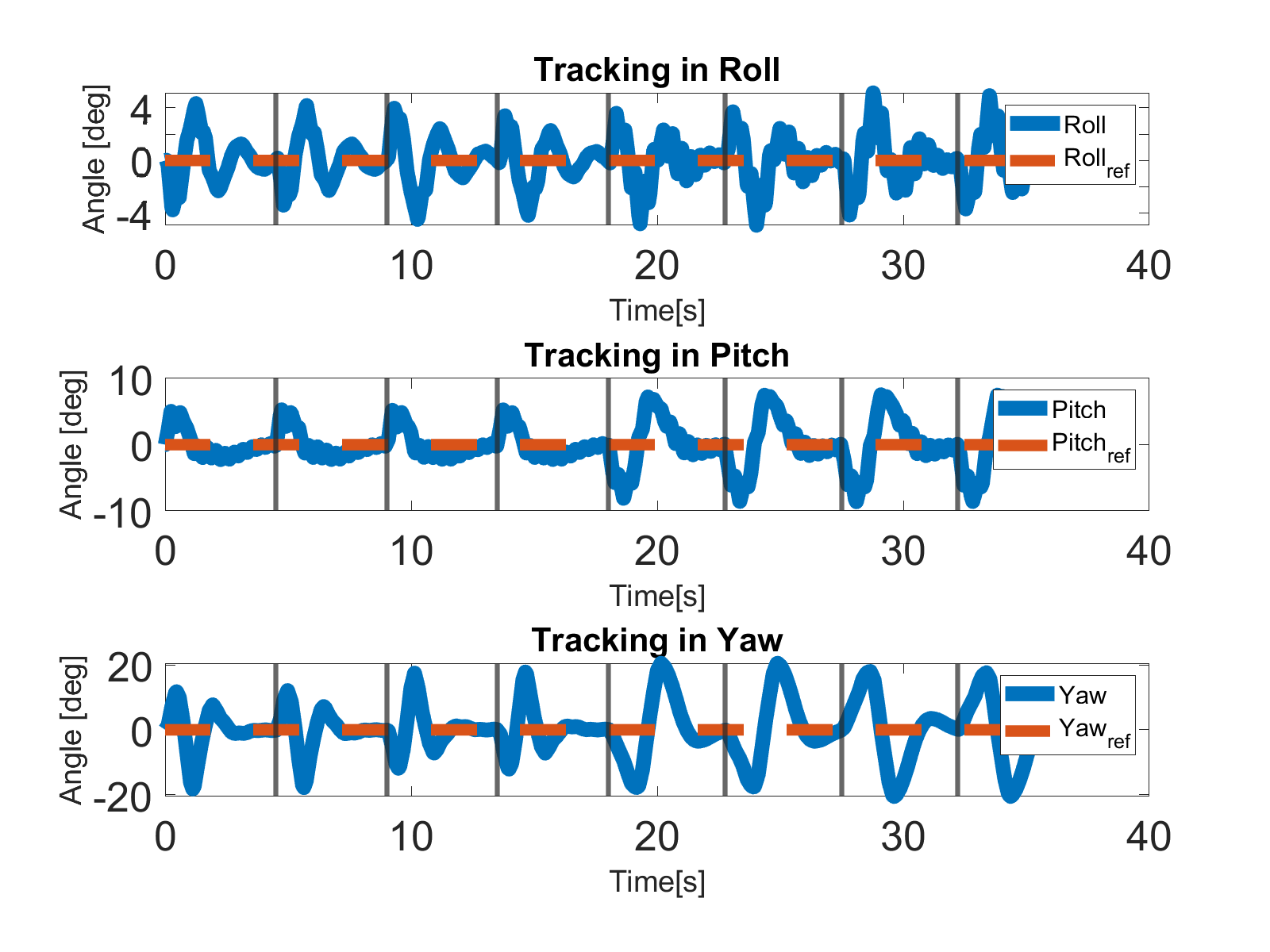}
  \caption{Attitude Variables Response}
  \label{fig::results_ind_tuned_att}
\end{subfigure}

%\caption{Trajectory Tracking with Independently Tuned Controllers}
\caption{Trajectory Tracking with Independent Tuning}
\label{fig::results_ind_tuned}
\end{figure}

\begin{figure}[thpb] 
\centering
\begin{subfigure}{.8\textwidth}
  \centering
  \includegraphics[width=\linewidth]{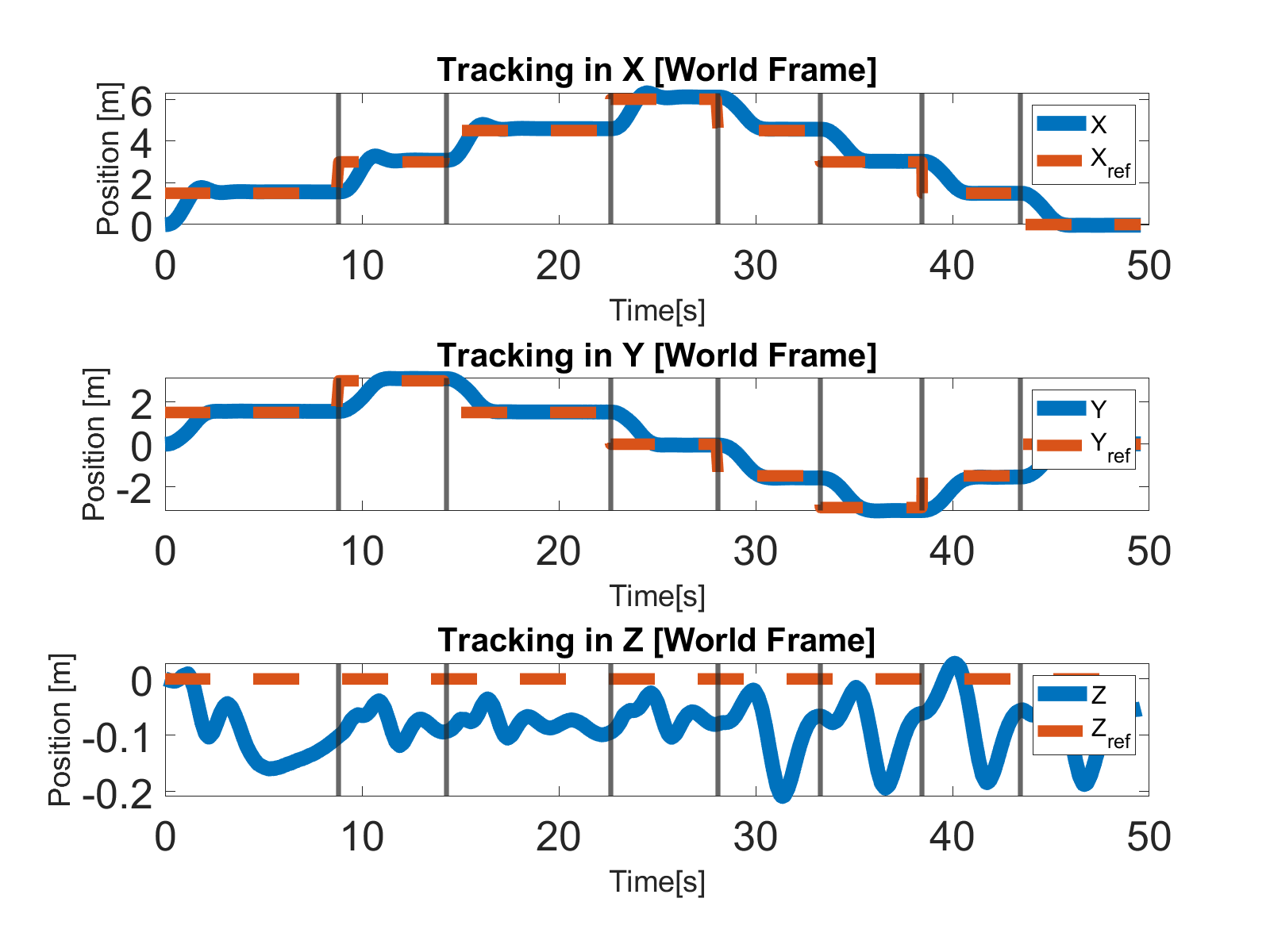}
  \caption{Position Variables Response}
  \label{fig::results_sim_tuned_pos}
\end{subfigure}
\begin{subfigure}{.8\textwidth}
  \centering
  \includegraphics[width=\linewidth]{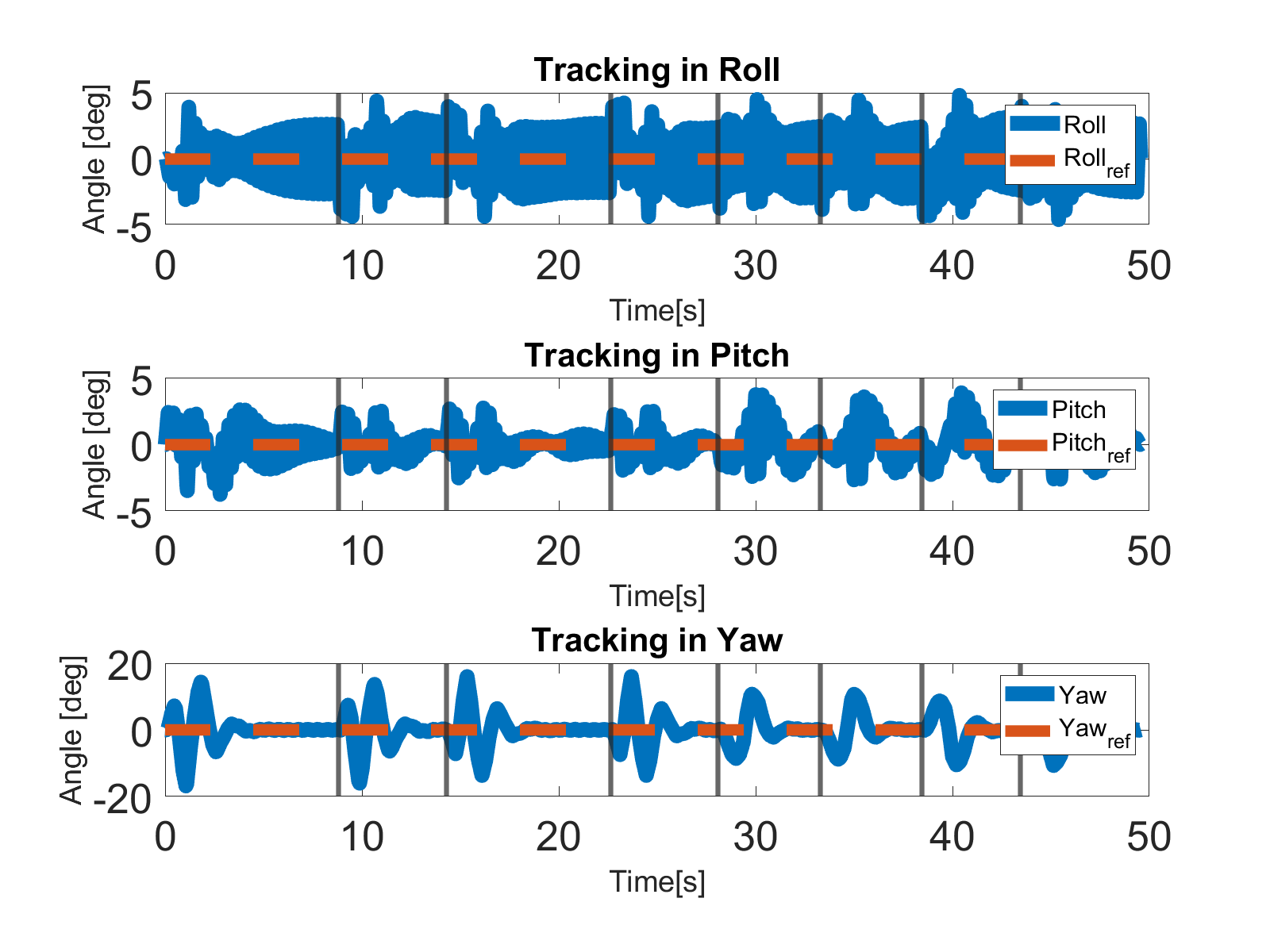}
  \caption{Attitude Variables Response}
  \label{fig::results_sim_tuned_att}
\end{subfigure}

\caption{Trajectory Tracking with Simultaneous Tuning}
%\caption{Trajectory Tracking with Simultaneously Tuned Controllers}
\label{fig::results_sim_tuned}
\end{figure}

Figs. \ref{fig::results_ind_tuned} and \ref{fig::results_sim_tuned} show the results of the trajectory tracking task performed with the individually tuned, and simultaneously tuned control configurations respectively (best case out of the 4 runs). Figs. \ref{fig::results_ind_tuned_att} and \ref{fig::results_sim_tuned_att} show the system response for the attitude variable tracking, where the goal for this particular trajectory is to keep the orientation of the vehicle (roll, pitch, and yaw) at a value of 0 while the XY trajectory is traversed. Figs. \ref{fig::results_ind_tuned_pos} and \ref{fig::results_sim_tuned_pos}, on the other hand, show the response for the position variables. In this case, the tracking task for the position in the $z$ axis is similar to the attitude variables, where the goal is to maintain an altitude of 0 throughout the whole trajectory. The $x$ and $y$ position plots show the sequence of reference way-point changes (marked in all sub-figures as vertical lines). Overall, the attitude variables show a more oscillatory behavior; which makes minimizing the tracking error for these variables a more difficult task. The trajectory completion time can be seen on the horizontal axis of all sub-figures. We can see that, for the best-case scenario, the individual tuning approach ($38.7s$ completion time) yields a control configuration with a better performance than the simultaneously tuned configuration ($49.6s$ completion time).
%% The response for the individually tuned control configuration is overall less oscillatory since with this approach, the controllers are each tuned to an error threshold, which guarantees a minimum individual performance for each controller. While the simultaneously tuned configuration cannot provide that guarantee. However, it has the advantage of being tuned for the specific task that is being evaluated.

% Individual

% Simultaneous

Fig. \ref{fig::results_xy_path} shows the results of the trajectory tracking task on the XY plane for the best case scenario (out of the 4 runs). The response for the individual tuning approach shows a lower tracking error in the XY plane compared to the simultaneously tuned control configuration. This can be explained by the exponential nature of the $\Upsilon_{eTxIAE}$ cost function, which heavily prioritizes episode completion time over tracking error. Depending on the application, a more expressive cost function could prioritize different aspects of the trajectory tracking task; however, such discussion is outside the scope of this paper.

% XY Plane

\subsection{Comparing Computational Complexity}\label{sec::results_comp_cost}
Table \ref{tbl::results_comp_cost} presents a detailed computational cost comparison with total time and sample complexity values. The individual tuning approach outperforms its simultaneous counterpart (as an average) by tuning controllers $86\%$ faster while requiring $36\%$ less iterations (system evaluations). The improvement in computational time is much larger since, as defined in Section \ref{sec::traj_task}, a square trajectory tracking episode (used in simultaneous tuning) can last up to 5 times longer than a step tracking episode (used in the individual tuning).

%Wilcoxon signed-rank test \cite{Hollander2015} was also used to compare computational complexity for both computational time (p-value = 0.0286) and number of samples (p-value = 0.2286).
% \begin{figure}[thbp]
%   \centering
%   \includegraphics[width=0.4\textwidth]{Images/Results_CompCost_Iter.png}
%   \caption{Comparing Sample Complexity}
%   \label{fig::results_comp_cost_sample}
% \end{figure}

\begin{table}[h] 
\caption{Computational Cost Comparison}
\newcolumntype{C}{>{\centering\arraybackslash}X}
\begin{center}
\begin{tabularx}{\textwidth}{C|CC}
\label{tbl::results_comp_cost}
\textbf{Comp. Cost} & \textbf{BO Individual} & \textbf{BO Simultaneous}\\
\hline
Time (hours)		& 8.41 (2.30)  & 61.46 (6.54) \\
\hline
\# Samples   & 613.25 (167.41) & 955.25 (89.50)\\
\end{tabularx}
\end{center}
\end{table}

%%%%%%%%%%%%%%%%%%%%%%%%%%%%%%%%%%%%%%%%%%
\section{Conclusions}
In this work, we have proposed a multi-stage optimization-based control tuning framework for MIMO systems with decentralized control. We apply this framework to decompose a control tuning task into optimization subtasks with smaller dimension search spaces in order to reduce computational complexity. We show that decomposing the parameter search space results in improved sample complexity and computational times when Bayesian Optimization is used to solve control the auto-tuning problem following the proposed framework. We empirically evaluate our hypotheses on an autonomous underwater vehicle benchmark with multiple PID control loops, achieving a 36\% and 86\% reduction in sample complexity and computational time respectively; with an improvement in trajectory tracking performance as well. We leave a set of open issues for future works, which include, controller performance objective design, local search space definition, robustness to system uncertainties, and a more detailed study on the relation between input-output variable decoupling and the performance of the proposed framework.

\bibliographystyle{vancouver}
\bibliography{ECC2024}

\begin{thebibliography}{10}

\bibitem{Mate2023}
Mate S, Pal P, Jaiswal A, Bhartiya S.
\newblock Simultaneous tuning of multiple PID controllers for multivariable systems using deep reinforcement learning.
\newblock Digital Chemical Engineering. 2023 12;9.

\bibitem{McClement2022}
McClement DG, Lawrence NP, Backström JU, Loewen PD, Forbes MG, Gopaluni RB.
\newblock Meta-reinforcement learning for the tuning of PI controllers: An offline approach.
\newblock Journal of Process Control. 2022 10;118:139-52.

\bibitem{Joseph2022}
Joseph SB, Dada EG, Abidemi A, Oyewola DO, Khammas BM. Metaheuristic algorithms for PID controller parameters tuning: review, approaches and open problems. Elsevier Ltd; 2022.

\bibitem{RodriguezMolina2020}
Rodríguez-Molina A, Mezura-Montes E, Villarreal-Cervantes MG, Aldape-Pérez M.
\newblock Multi-objective meta-heuristic optimization in intelligent control: A survey on the controller tuning problem.
\newblock Applied Soft Computing Journal. 2020 8;93.

\bibitem{Stenger2022bench}
Stenger D, Abel D.
\newblock Benchmark of Bayesian Optimization and Metaheuristics for Control Engineering Tuning Problems with Crash Constraints; 2022.

\bibitem{Stenger2022joint}
Stenger D, Nitsch M, Abel D.
\newblock Joint Constrained Bayesian Optimization of Planning, Guidance, Control, and State Estimation of an Autonomous Underwater Vehicle.
\newblock In: 2022 European Control Conference, ECC 2022. IEEE; 2022. p. 1982-7.

\bibitem{Rohr2024}
von Rohr A, Stenger D, Scheurenberg D, Trimpe S.
\newblock Local Bayesian optimization for controller tuning with crash constraints.
\newblock At-Automatisierungstechnik. 2024 4;72:281-92.

\bibitem{Nitsch2023}
Nitsch M, Stenger D, Abel D.
\newblock Automated Tuning of Nonlinear Kalman Filters for Optimal Trajectory Tracking Performance of AUVs.
\newblock In: IFAC-PapersOnLine. vol.~56. Elsevier B.V.; 2023. p. 11608-14.

\bibitem{Paulson2023}
Paulson JA, Sorourifar F, Mesbah A.
\newblock A Tutorial on Derivative-Free Policy Learning Methods for Interpretable Controller Representations.
\newblock In: Proceedings of the American Control Conference. vol. 2023-May. IEEE; 2023. p. 1295-306.

\bibitem{Somefun2021}
Somefun OA, Akingbade K, Dahunsi F. The dilemma of PID tuning. Elsevier Ltd; 2021.

\bibitem{Lawrence2022}
Lawrence NP, Forbes MG, Loewen PD, McClement DG, Backström JU, Gopaluni RB.
\newblock Deep reinforcement learning with shallow controllers: An experimental application to PID tuning.
\newblock Control Engineering Practice. 2022 4;121.

\bibitem{Kumar2021}
Kumar AR, Ramadge PJ.
\newblock DiffLoop: Tuning PID controllers by differentiating through the feedback loop.
\newblock In: 2021 55th Annual Conference on Information Sciences and Systems, CISS 2021. Institute of Electrical and Electronics Engineers Inc.; 2021. .

\bibitem{Pandey2022}
Pandey SK, Dey J, Banerjee S.
\newblock Generalized discrete decoupling and control of MIMO systems.
\newblock Asian Journal of Control. 2022 11;24:3326-44.

\bibitem{Sahrir2022}
Sahrir NH, Basri MAM.
\newblock Modelling and Manual Tuning PID Control of Quadcopter.
\newblock In: Lecture Notes in Electrical Engineering. vol. 921 LNEE. Springer Science and Business Media Deutschland GmbH; 2022. p. 346-57.

\bibitem{Liu2019survey}
Liu L, Tian S, Xue D, Zhang T, Chen YQ, Zhang S.
\newblock A Review of Industrial MIMO Decoupling Control.
\newblock International Journal of Control, Automation and Systems. 2019 5;17:1246-54.

\bibitem{Agrawal2020}
Agrawal A, Barratt S, Boyd S, Stellato B, Bayen A, Jadbabaie A, et~al.
\newblock Learning Convex Optimization Control Policies.
\newblock In: Proceedings of Machine Learning Research. vol. 120; 2020. p. 1-13.
\newblock Available from: \url{https://web.stanford.edu/}.

\bibitem{Trimpe2014}
Trimpe S, Millane A, Doessegger S, D'andrea R.
\newblock A Self-Tuning LQR Approach Demonstrated on an Inverted Pendulum; 2014.

\bibitem{Carlucho2020}
Carlucho I, Paula MD, Acosta GG.
\newblock An adaptive deep reinforcement learning approach for MIMO PID control of mobile robots.
\newblock ISA Transactions. 2020 7;102:280-94.

\bibitem{Cheng2023}
Cheng S, Song L, Kim M, Wang S, Hovakimyan N.
\newblock DiffTune+: Hyperparameter-Free Auto-Tuning using Auto-Differentiation. 2023 5.
\newblock Available from: \url{http://arxiv.org/abs/2212.03194}.

\bibitem{Chang2014}
Chang WD, Chen CY.
\newblock PID controller design for MIMO processes using improved particle swarm optimization.
\newblock Circuits, Systems, and Signal Processing. 2014;33:1473-90.

\bibitem{Yadav2020}
Yadav S, Mishra A, Nagar SK.
\newblock Tuning of parameters of PID controller using Grey Wolf Optimizer.
\newblock In: Proceedings of the International Conference on Advances in Electronics, Electrical and Computational Intelligence (ICAEEC) 2019; 2020. .

\bibitem{Zhou2019moo}
Zhou X, Zhang X.
\newblock Multi-objective-optimization-based control parameters auto-tuning for aerial manipulators.
\newblock International Journal of Advanced Robotic Systems. 2019 1;16.

\bibitem{Hekimoglu2019}
Hekimoglu B.
\newblock Optimal Tuning of Fractional Order PID Controller for DC Motor Speed Control via Chaotic Atom Search Optimization Algorithm.
\newblock IEEE Access. 2019;7:38100-14.

\bibitem{Mulders2020}
Mulders SP, Pamososuryo AK, Wingerden JWV.
\newblock Efficient tuning of Individual Pitch Control: A Bayesian Optimization Machine Learning approach.
\newblock In: Journal of Physics: Conference Series. vol. 1618. IOP Publishing Ltd; 2020. .

\bibitem{Kavas2024}
Kavas B, Balta EC, Tucker MR, Krishnadas R, Rupenyan A, Lygeros J, et~al.
\newblock In-situ Controller Autotuning by Bayesian Optimization for Closed-loop Feedback Control of Laser Powder Bed Fusion Process. 2024 6.
\newblock Available from: \url{http://arxiv.org/abs/2406.19096}.

\bibitem{Stenger2023thesis}
Stenger D.
\newblock Automatic Tuning of Control Engineering Algorithms with Bayesian Optimization. 2023.

\bibitem{Jones1998}
Jones DR, Schonlau M, Welch WJ.
\newblock Efficient Global Optimization of Expensive Black-Box Functions.
\newblock Journal of Global Optimization. 1998;13:455-92.

\bibitem{Kushner1964}
Kushner HJ.
\newblock A New Method of Locating the Maximum Point of an Arbitrary Multipeak Curve in the Presence of Noise.
\newblock Journal of Basic Engineering. 1964.

\bibitem{Mockus1975}
Mockus J.
\newblock On bayesian methods for seeking the extremum.
\newblock In: Marchuk GI, editor. Optimization Techniques IFIP Technical Conference Novosibirsk, July 1–7, 1974. Springer Berlin Heidelberg; 1975. p. 400-4.

\bibitem{Rasmussen2006}
Rasmussen CE, Williams CKI.
\newblock Gaussian Processes for Machine Learning.
\newblock USA:MIT Press; 2006.

\bibitem{Shahriari2016}
Shahriari B, Swersky K, Wang Z, Adams RP, de~Freitas N.
\newblock Taking the Human Out of the Loop: A Review of Bayesian Optimization.
\newblock Proceedings of the IEEE. 2016 1;104:148-75.

\bibitem{Rana2017}
Rana S, Li C, Gupta S, Nguyen V, Venkatesh S.
\newblock High dimensional Bayesian optimization with elastic Gaussian Process.
\newblock In: International conference on machine learning; 2017. p. 2883-91.

\bibitem{Garnett2014}
Garnett R, Osborne MA, Hennig P.
\newblock Active learning of linear embeddings for Gaussian processes.
\newblock In: Conference on uncertainty in artificial intelligence; 2014. p. 230-9.

\bibitem{Cerqueira2022}
de~Cerqueira~Gava PD, Júnior CLN, de~França~Silva JRB, Adabo GJ.
\newblock Simu2VITA: A General Purpose Underwater Vehicle Simulator.
\newblock Sensors (Basel, Switzerland). 2022 4;22.

\bibitem{Fossen2011}
Fossen TI.
\newblock Handbook of marine craft hydrodynamics and motion control.
\newblock Wiley; 2011.

\end{thebibliography}

\end{document}